\documentclass[10pt]{article}
\usepackage{latexsym}
\usepackage{amssymb}
\usepackage{amsmath}
\usepackage{amscd}
\usepackage{amsthm}
\usepackage[left=2cm,top=2.5cm,right=2.5cm,bottom=1.5cm]{geometry}
\usepackage[dvips]{graphicx}
\usepackage{epstopdf}
\usepackage{hyperref}
\begin{document}
\begin{center}
\large{\bf{Phantom Instability of Viscous Dark Energy in Anisotropic Space-Time}} \\
\vspace{10mm}
\normalsize{Hassan Amirhashchi$^{1,2}$}\\
\vspace{5mm}
\normalsize{$^1$Young Researchers Club, Mahshahr Branch, Islamic Azad University, Mahshahr, Iran \\
$^2$Laboratory of Computational
Sciences and Mathematical Physics, Institute for Mathematical
Research, University Putra Malaysia, 43400 Serdang, Selangor D.E., Malaysia \\
$^1$E-mail:h.amirhashchi@mahshahriau.ac.ir}  \\
\end{center}
\vspace{10mm}
\begin{abstract}
Phantom dark energy is a proposal that explains the current
observations that mildly favor the equation of state of dark energy
$\omega^{de}$ crossing $-1$ at $68\%$ confidence level. However,
phantom fields are generally ruled out by ultraviolet quantum
instabilities. To overcome this discrepancy, in this paper we
propose a mechanism to show that how the presence of bulk
viscosity in the cosmic fluid can temporarily drive the fluid into
the phantom region ($\omega<-1$). As time is going on, phantom
decays and ultimately $\omega^{de}$ approaches to $-1$. Then we
show these quintessence and phantom descriptions of non-viscous
and viscous dark energy and reconstruct the potential of these two
scalar fields. Also a diagnostic for these models are performed by
using the statefinder pairs $\{s,r\}$. All results are obtained in
an anisotropic space-time which is a generalization of FLRW
universe.
\end{abstract}
\section{Introduction}
It is a very well known fact that our universe is experiencing an
accelerating expansion at the present time
(Perlmutter et al. 1999; Riess et al. 1998, 2001; Tonry et al. 2003; Tegmark et al. 2004). It is believed that an exotic form of
energy with negative pressure called dark energy is responsible
for the current observed accelerating expansion of the universe
(Tegmark et al. 2004; Bennet et al. 2003; Spergel et al. 2003; Abazajian et al. 2004). According to the recent observations we
live in a nearly spatially flat Universe composed of approximately
$4\%$ baryonic matter, $22\%$ dark matter and $74\%$ dark energy.
However, the observational data are far from being complete. It is
not even known what is the current value of the dark energy
effective equation of state (EoS) parameter $\omega^{(de)} =
p^{(de)}/\rho^{(de)}$ which lies close to $-1$: it could be equal
to $-1$ (standard $\Lambda$CDM cosmology), a little bit upper than
$-1$ (the quintessence dark energy) or less than $-1$ (phantom
dark energy). One of the main candidate for dark energy is
cosmological constant $\Lambda$, which has pressure $p^{(de)} = -
\rho^{(de)}$. Although, cosmological constant can explain the
current acceleration phase of universe, it would suffer from many
serious theoretical problems, such as the fine-tuning and the
coincidence problems. Another candidate for dark energy is
provided by introducing scalar fields. An important class of
scalar fields are known as ``quintessence" with
$-\frac{1}{3}>\omega>-1$ (Ratra and Peebles. 1988; Wetterich 1988; Turner and white 1997; Caldwell et al. 1998; Liddle and Scherrer 1999; Steinhardt et al. 1999) in which the
scalar field mimics the perfect fluid and hence could lead to a
solution for coincidence problem. However, quintessence scenario
of dark energy is not in accurate consistent with recent
observations as $\omega<-1$ has been favored by recent
observations (Knop et al. 2003; Riess et al. 2004; Alam et al. 2004; Hannestad and E. Mortsell 2004). To get $\omega<-1$, a new
class of scalar field models with negative kinetic energy, known
as ``phantom field" models have been suggested (Caldwell 2002).
Nevertheless, in this case the universe shows some very strange
properties (Carroll et al. 2003; Cline et al. 2004; Buniy and Hsu 2006; Buniy et al. 2006). For example, since the
energy density of phantom field is unbounded from below, the
vacuum becomes unstable against the production of positive energy
fields hence these fields are generally ruled out by ultraviolet
quantum instabilities (Carroll et al. 2003). Another problem is the future
finite singularity called Big Rip (Caldwell et al. 2003) which leads to the
occurrence of negative entropy (Brevik et al. 2004). Therefore, on the one
hand observations mildly favors models with $\omega$ crossing $-1$
near the past and on another, models with $\omega<-1$ are unstable
from theoretical point of view. In this paper we suggest a simple
mechanism to overcome this discrepancy by introducing bulk
viscosity in the cosmic fluid. First, viscosity causes dark energy
which is varying in quintessence to pass the phantom divided line
(PDL) and drop it to phantom region. Next, since viscosity is a
decreasing function of time, it will die out and $\omega$ will
leave phantom region and tend to $-1$ at late time. Hence the
problem of future
singularity (big rip) will never occur in this scenario.\\

It has been shown in refs (McInnes 2002; Barrow 2004) that, an ideal cosmic
fluid, i.e. non-viscous, give raise to the occurrence of a
singularity of the universe in the far future called big rip. The
singularity problem can be modified or soften via following two
methods. The first is the effect of quantum corrections due to the
conformal anomaly (Brevik and Odintsov 1999; Nojiri and Odintsov 2003, 2004) and second, is to
consider the bulk viscosity of the cosmic fluid (for example see
(Misner 1968; Padmanabhan and Chitre 1987; Brevik and Hallanger 2004). The viscosity theory of
relativistic fluids was first suggested by Eckart, Landau and
Lifshitz (Eckart 1940; Landau and Lifshitz 1987). The introduction of viscosity into
cosmology has been investigated from different view points
(Gr$\o$n 1990; Barrow 1986; Zimdahl 1996; Maartens 1996). The astrophysical observations also
indicate some evidences that cosmic media is not a perfect fluid
(Jaffe et al. 2005), and the viscosity effect could be concerned in the
evolution of the universe (Brevik and Gorbunova 2005; Brevik et al. 2005; Cataldo et al. 2005). It was also
argued that a viscous pressure can play the role of an agent that
drives the present acceleration of the Universe
(Zimdahl et al. 2001; Balakin et al. 2003). The possibility of a viscosity dominated late
epoch of the Universe with accelerated expansion was already
mentioned by Padmanabhan and Chitre (Padmanabhan and Chitre 1987). Brevik and
Gorbunova (2005), Oliver et al (2011), Chen et al
(2011), Jamil and Farooq (2010), Cai et al (2010), Setare (2007a, 2007b, 2007c), Setare et al (2007), Setare and Saridakis (2009), Setare et al (2009), Amirhashchi et al (2011 a, 2011b, 2011 c, 2013), Pradhan et al (2011a, 2011b, 2011c), Saha et al (2012), and Sheykhi and
Setare (2010) have studied viscous and non-viscous dark energy models in
different contexts. Recently, viscous dark energy and generalized
second law of thermodynamics has been studied by Setare and
Sheykhi (2010).\\

To be general, we use generalized FLRW equations by considering an
anisotropic metric as the line-element of the universe. The reason
for this choice of metric is behind the fact that because of high
symmetry, FLRW models are infinitely improbable in the space of
all possible cosmologies. The high symmetry involved in FLRW
models requires a very high degree of fine tuning of initial
conditions which is extraordinary improbable. Moreover, we can
always ask that does the universe necessarily have the same
symmetries on very large scales outside the particle horizon or at
early times?\\
The plan of our paper is as follows: In section $2$ we give the
metric and field equations. In section $3$ we drive the
generalized FLRW equations by solving the field equations of
section $2$. The general form of non-viscous and viscous dark
energy equation of state parameter EoS are given in section $4$. We
suggest a correspondence between the non-viscous and viscous dark
energy scenario and the quintessence and phantom dark energy model
in section $5$. In section $6$, a statefinder diagnostic has been
presented. In section $7$ we apply our general results to a toy
model in order to test the proposed mechanism. Our results are
summarized in section $8$.
\section{The Metric and Field  Equations}
We consider the Bianchi type I space-time in the orthogonal form
as
\begin{equation}
\label{eq1}
ds^{2} = -dt^{2} + A^{2}(t)dx^{2}+B^{2}(t)dy^{2}+C^{2}(t)dz^{2},
\end{equation}
where $A(t), B(t)$ and $C(t)$ are functions of time only. \\

The Einstein's field equations ( in gravitational units $8\pi G = c = 1 $) read as
\begin{equation}
\label{eq2} R^{i}_{j} - \frac{1}{2} R g^{i}_{j} = T^{(m)i}_{j} +
T^{(de)i}_{j},
\end{equation}
where $T^{(m)i}_{j}$ and $T^{(de)i}_{j}$ are the energy momentum tensors of barotropic matter and dark energy,
respectively. These are given by
\[
  T^{(m)i}_{j} = \mbox{diag}[-\rho^{(m)}, p^{(m)}, p^{(m)}, p^{(m)}],
\]
\begin{equation}
\label{eq3} ~ ~ ~ ~ ~ ~ ~ ~  = \mbox{diag}[-1, \omega^{(m)}, \omega^{(m)}, \omega^{(m)}]\rho^{m},
\end{equation}
and
\[
 T^{(de)i}_{j} = \mbox{diag}[-\rho^{(de)}, p^{(de)}, p^{(de)}, p^{(de)}],
\]
\begin{equation}
\label{eq4} ~ ~ ~ ~ ~ ~ ~ ~ ~ ~ ~ ~ ~ ~ = \mbox{diag}[-1, \omega^{(de)}, \omega^{(de)},
\omega^{(de)}]\rho^{(de)},
\end{equation}
where $\rho^{(m)}$ and $p^{(m)}$ are, respectively the energy density and pressure of the perfect fluid
component or matter while $\omega^{(m)} = p^{(m)}/\rho{(m)}$ is its EoS parameter. Similarly,
$\rho^{(de)}$ and $p^{(de)}$ are, respectively the energy density and pressure of the DE component while
$\omega^{(de)} = p^{(de)}/\rho^{(de)}$ is the corresponding EoS parameter. We assume the four velocity vector
$u^{i} = (1, 0, 0, 0)$ satisfying $u^{i}u_{j} = -1$. \\

In a co-moving coordinate system ($u^{i} = \delta^{i}_{0}$), Einstein's field equations (\ref{eq2}) with
(\ref{eq3}) and (\ref{eq4}) for B-I metric (\ref{eq1}) subsequently lead to the following system of equations:
\begin{equation}
\label{eq5} \frac{\ddot{B}}{B}+\frac{\ddot{C}}{C}+\frac{\dot{B}\dot{C}}{BC}=-\omega^{m}\rho^{m}-\omega^{de}\rho^{de},
\end{equation}
\begin{equation}
\label{eq6} \frac{\ddot{A}}{A}+\frac{\ddot{C}}{C}+\frac{\dot{A}\dot{C}}{AC}=-\omega^{m}\rho^{m}-\omega^{de}\rho^{de},
\end{equation}
\begin{equation}
\label{eq7}\frac{\ddot{A}}{A}+\frac{\ddot{B}}{B}+\frac{\dot{A}\dot{B}}{AB}=-\omega^{m}\rho^{m}-\omega^{de}\rho^{de},
\end{equation}
\begin{equation}
\label{eq8} \frac{\dot{A}\dot{B}}{AB}+\frac{\dot{A}\dot{C}}{AC}+\frac{\dot{B}\dot{C}}{BC}=\rho^{m}+\rho^{de}.
\end{equation}
If we consider $a=(ABC)^{\frac{1}{3}}$ as the average scale factor of Bianchi type I model, then
the generalized mean Hubble's parameter $H$ defines as
\begin{equation}
\label{eq9} H = \frac{\dot{a}}{a} = \frac{1}{3}\left(\frac{\dot{A}}{A} + \frac{\dot{B}}{B} + \frac{\dot{C}}{C}\right).
\end{equation}

The Bianchi identity $G^{;j}_{ij} = 0$ leads to  $T^{;j}_{ij} = 0$. Therefore, the continuity equation
for dark energy and baryonic matter can be written as
\begin{equation}
\label{eq10} \dot{\rho}^{m} + 3H(1 + \omega^{m})\rho^{m} + \dot{\rho}^{de} + 3H(1 + \omega^{de})\rho^{de} = 0.
\end{equation}
\section{Friedmann-Like Equations}
In this section, we derive the general solution for
the Einstein's field equations (\ref{eq5})-(\ref{eq8}).\\

Subtracting Eq. (\ref{eq5}) from Eq. (\ref{eq6}), Eq. (\ref{eq6})
from Eq. (\ref{eq7}), and Eq. (\ref{eq5}) from Eq. (\ref{eq7}) we
obtain
\begin{equation}
\label{eq11} \frac{\ddot{A}}{A} - \frac{\ddot{B}}{B} +
\frac{\dot{C}}{C}\left(\frac{\dot{A}}{A} -
\frac{\dot{B}}{B}\right) = 0,
\end{equation}

\begin{equation}
\label{eq12} \frac{\ddot{B}}{B} - \frac{\ddot{C}}{C} +
\frac{\dot{A}}{A}\left(\frac{\dot{B}}{B} -
\frac{\dot{C}}{C}\right) = 0,
\end{equation}
and
\begin{equation}
\label{eq13} \frac{\ddot{A}}{A} - \frac{\ddot{C}}{C} +
\frac{\dot{B}}{B}\left(\frac{\dot{A}}{A} -
\frac{\dot{C}}{C}\right) = 0.
\end{equation}
First integral of Eqs. (\ref{eq11}), (\ref{eq12}) and (\ref{eq13})
leads to
\begin{equation}
\label{eq14}
\frac{\dot{A}}{A}-\frac{\dot{B}}{B}=\frac{k_{1}}{ABC},
\end{equation}
and
\begin{equation}
\label{eq15}
\frac{\dot{B}}{B}-\frac{\dot{C}}{C}=\frac{k_{2}}{ABC},
\end{equation}
\begin{equation}
\label{eq16}
\frac{\dot{A}}{A}-\frac{\dot{C}}{C}=\frac{k_{3}}{ABC},
\end{equation}
where $k_{1}$, $k_{2}$ and $k_{3}$ are constants of integration.
By taking integral from Eqs. (\ref{eq14}), (\ref{eq15}) and
(\ref{eq16}) we get
\begin{equation}
\label{eq17} \frac{\dot{A}}{B}=d_{1}exp[k_{1}\int(ABC)^{-1}dt],
\end{equation}

\begin{equation}
\label{eq18} \frac{\dot{B}}{C}=d_{2}exp[k_{2}\int(ABC)^{-1}dt],
\end{equation}
and
\begin{equation}
\label{eq19} \frac{\dot{A}}{C}=d_{3}exp[k_{3}\int(ABC)^{-1}dt]
\end{equation}
where, $d_{1}, d_{2}$ and $d_{3}$ are constants of integration.\\
Now, we can find all metric potentials from Eqs. (\ref{eq17}),
(\ref{eq19}) as follow
\begin{equation}
\label{eq20} A(t)=a_{1}a~ exp(b_{1}\int a^{-3}dt),
\end{equation}
\begin{equation}
\label{eq21} B(t)=a_{2}a~ exp(b_{2}\int a^{-3}dt),
\end{equation}
and
\begin{equation}
\label{eq22} C(t)=a_{3}a~ exp(b_{3}\int a^{-3}dt).
\end{equation}
Here
\[
a_{1}=(d_{1}d_{2})^{\frac{1}{3}},~~~~~a_{2}=(d_{1}^{-1}d_{3})^{\frac{1}{3}},~~~~~a_{3}=(d_{2}d_{3})^{-\frac{1}{3}}
,~~~~~b_{1}=\frac{k_{1}+k_{2}}{3},~~~~~b_{2}=\frac{k_{3}-k_{1}}{3},~~~~~b_{3}=-\frac{k_{2}+k_{3}}{3},
\]
where
\[
a_{1}a_{2}a_{3}=1,~~~~~~~b_{1}+b_{2}+b_{3}=0.
\]

Therefore, one can write the general form of Bianchi type I metric
as
\begin{equation}
\label{eq23} ds^{2}=-dt^{2}+a^{2}\left[a_{1}^{2} e^{2b_{1}\int
a^{-3}dt}dx^{2}+a_{2}^{2} e^{2b_{2}\int a^{-3}dt}dy^{2}+a_{3}^{2}
e^{2b_{3}\int a^{-3}dt}dz^{2}\right].
\end{equation}

Using eqs. (\ref{eq20})-(\ref{eq22}) in eqs.
(\ref{eq5})-(\ref{eq8}) we can write the analogue of the Friedmann
equation as
\begin{equation}
\label{eq24} \left(\frac{\dot{a}}{a}\right)^{2}=\frac{\rho}{3}+ K
a^{-6},
\end{equation}
and
\begin{equation}
\label{eq25}
2\left(\frac{\ddot{a}}{a}\right)=-\frac{1}{3}(\rho+3p).
\end{equation}
Here $\rho = \rho^{m} + \rho^{de}$, $p = p^{m} + p^{de}$ and
$K=b_{1}b_{2}+b_{1}b_{3}+b_{2}b_{3}$. Note that $K$ denotes the
deviation from isotropy e.g. $K=0$ represents flat FLRW universe.
Thus, when the universe is sufficiently large, almost at the present time, the space-time
(\ref{eq1}) behaves like a flat FLRW universe.
\section{Dark Energy Equation of State}
In this section we obtain the general form of the equation of
state (EoS) for the viscous and non viscous dark energy (DE)
$\omega^{de}$ in Bianchi type I space-time when there is no
interaction between dark energy and Cold Dark Matter(CDM) with
$\omega_{m}=0$. In this case the conservation equation
(\ref{eq10}) for dark and barotropic fluids can be written
separately as
\begin{equation}
\label{eq26} \dot{\rho}^{de} + 3H(1 + \omega^{de})\rho^{de} = 0,
\end{equation}
and
\begin{equation}
\label{eq27} \dot{\rho}^{m} + 3H\rho^{m}=0.
\end{equation}
Eq.(\ref{eq27})leads to
\begin{equation}
\label{eq28} \rho^{m}=\rho_{0}^{m}a^{-3}.
\end{equation}
Using eqs. (\ref{eq24}), (\ref{eq28}) in eqs. (\ref{eq7}),
(\ref{eq8}) we obtain the energy density and pressure of dark
fluid as
\begin{equation}
\label{eq29} \rho^{de}=3H^{2}-3Ka^{-6}-\rho_{0}^{m}a^{-3}
\end{equation}
and
\begin{equation}
\label{eq30} p^{de}=-2\frac{\ddot{a}}{a}-H^{2}-La^{-6},
\end{equation}
respectively. Therefore, the equation of state parameter (EoS) of
DE in it's general form is given by
\begin{equation}
\label{eq31}
\omega^{de}_{pf}=\frac{p^{de}}{\rho^{de}}=\frac{2q-1-La^{-6}H^{-2}}{3+3La^{-6}H^{-2}-3\Omega_{0}^{m}a^{-3}},
\end{equation}
where $q=-\frac{\ddot{a}}{aH^{2}}$ is the deceleration parameter,
$\Omega_{0}^{m}$ is the current value of matter density and
$L=b_{2}^{2}+b_{3}^{2}+b_{2}b_{3}$ is a positive constant (Note
that $K+L=0$).\\

 From eq. (\ref{eq31}) we see that at present time (i.e $L=0,~
 q=-0.55,~
\Omega_{0}^{m}=0.24,~ a=1 $), approximately,
$\omega^{de}_{pf}=-0.92$. At late time, EoS parameter is given by
\begin{equation}
\label{eq32} \omega^{de}_{pf}\sim \frac{2q-1}{3},
\end{equation}
here subscript `$pf$' refers to ``perfect fluid".\\

According to the observations deceleration parameter is restricted
as $-1\leq q<0$. Therefore, from eq. (\ref{eq32}) we observe that
 at the best approximation the minimum value of $\omega^{de}_{pf}$
is $-1$ i.e  EoS of non-viscous DE can not cross phantom divided
line (PDL). In another word, non-viscous dark energy can be
described by quintessence ($\omega^{de}>-1$) rather than phantom
($\omega^{de}<-1$) field. In other hand models with $\omega^{de}$
crossing $-1$ near the past have been mildly favored by the
analysis on the nature of dark energy from recent observations
(for example see (Astier et al. 2006)). SNe Ia alone favors a $\omega$
larger than $-1$ in the recent past and less than $-1$ today,
regardless of wether using the prior of a flat universe
(Alam et al. 2004; Astier et al. 2006) or not (Dicus and Repko 2004). In what follows we show
that the possibility of crossing PDL will be available in
a universe dominated by viscous DE instead of perfect DE.\\

In Eckart's theory (Eckart 1940) a viscous dark energy EoS is
specified by
\begin{equation}
\label{eq33} {p}^{de}_{vf} = p^{de}_{pf} + \Pi.
\end{equation}
Here $\Pi = -\xi(\rho^{de})u^{i}_{;i}$ is the viscous pressure and
$H = \frac{u^{i}_{;i}}{3}$ is the Hubble's parameter and subscript
`$vf$' referees to ``viscous fluid". On thermodynamical grounds,
in conventional physics $\xi$ has to be positive. This is a
consequence of the positive sign of the entropy change in an
irreversible process (Nojiri and Odintsov 2003). In general,
$\xi(\rho^{de})=\xi_{0}(\rho^{de})^{\tau}$, where $\xi_{0}>0$ and
$\tau$ are constant parameters. Note that, here we have to assume $\tau>0$
since for negative $\tau$ this form of bulk viscosity does not allow our models to cross
PDL. A power-law expansion for the scale factor can be achieved for $\tau=\frac{1}{2}$ (Barrow 1987, 1988).
It has been shown by Goliath and Ellis (1999) that some Bianchi
models isotropise due to inflation.\\

Substituting eq. (\ref{eq33}) in eq. (\ref{eq31}) by considering
the above description we obtain the EoS parameter of viscous DE as
\begin{equation}
\label{eq34} \omega^{de}_{vf}
=\frac{p^{de}}{\rho^{de}}+\frac{\Pi}{\rho^{de}}=\frac{2q-1-La^{-6}H^{-2}}{3+3La^{-6}H^{-2}
-3\Omega_{0}^{m}a^{-3}}-3\xi_{0}\frac{H^{1-2\alpha}}{(3\Omega^{de})^{\alpha}},
\end{equation}
where $\Omega^{de}=\frac{\rho^{de}}{3H^{2}}$ and $\alpha=1-\tau$. \\
From eq. (\ref{eq34}) we observe that the EoS of viscous DE at
present time (i.e $L=0,~ q=-0.55,~H_{0}=70,~
\Omega_{0}^{m}=0.24,~\Omega_{0}^{de}=0.76,~ a=1 $), approximately
is
\begin{equation}
\label{eq35} \omega^{de}_{vf}\sim
-0.92-\frac{213\xi_{0}}{(12501.68)^{\alpha}},
\end{equation}
which clearly cross the PDL for appropriate values of $\alpha$ and
$\xi_{0}$. As mentioned before, phantom fields are generally
plagued by ultraviolet quantum instabilities. Naively, any phantom
model with $\omega^{de}<-1$ should decay to $\omega^{de}=-1$ at
late time. As mentioned in  (Carroll et al. 2003), this ensures that there
is no future singularity (Big Rip); rather, the universe
eventually settles into a de Sitter phase. Here we highlight since
$\xi(\rho^{de})=\xi_{0}(\rho^{de})^{\tau}$, and $\rho^{de}$ is a
decreasing function of time in an expanding universe we conclude
that the bulk viscosity dies out as time goes on and viscous
phantom DE is an unstable state (as expected) and EoS of DE tends
to $-1$ at late time (de-Sitter Universe).
\section{Correspondence Between Dark Energy And Scalar Fields}
It is believed that the current accelerated expansion is driven by
a dynamical scalar field $\phi$ with potential $V(\phi)$. These
models introduce a scalar field $\phi$ that is minimally coupled
to gravity. As it is shown in previous section, one can generate
quintessence and phantom fields from non-viscous and viscous
fluids in an anisotropic universe respectively.\\
Quintessence and phantom fields are generally given by the action
\begin{equation}
\label{eq36} S=\int
d^{4}x\sqrt{-g}\left[-\frac{1}{2}\epsilon(\nabla
\phi)^{2}-V(\phi)\right].
\end{equation}
The energy density and pressure of scalar field (DE) are given by
\begin{equation}
\label{eq37} \rho_{\phi}=\frac{1}{2}\epsilon
\dot{\phi}^{2}+V(\phi)
\end{equation}
and
\begin{equation}
\label{eq38}p_{\phi}=\frac{1}{2}\epsilon \dot{\phi}^{2}-V(\phi),
\end{equation}
where $\epsilon=\pm 1$. $\epsilon=1$ is referred to as
quintessence whereas $\epsilon=-1$ is referred to as phantom. From
eqs. (\ref{eq29}), (\ref{eq30}) and eqs. (\ref{eq37}),
(\ref{eq38}) we find the general form of the scalar field $\phi$
and potential $V(\phi)$ as
\begin{equation}
\label{eq39} \dot{\phi}^{2}=2\epsilon
\left[H^{2}(1+q)+La^{-6}-\frac{3}{2}H^{2}\Omega_{0}^{m}a^{-3}-\frac{\xi_{0}}{2}\sqrt{3\Omega^{de}}\right],
\end{equation}
and
\begin{equation}
\label{eq40}
V(\phi)=2\left[H^{2}(1-q)-4La^{-6}-\frac{3}{2}H^{2}\Omega_{0}^{m}a^{-3}+\frac{\xi_{0}}{2}\sqrt{3\Omega^{de}}\right].
\end{equation}
Note that putting $\xi_{0}=0$ and $\epsilon=1$ in eqs.
(\ref{eq39}), (\ref{eq40}) we get the scalar field and potential
of quintessence. Also for sufficiently large time, the asymptotic
behavior of $\phi$ and $V(\phi)$ is given by
\begin{equation}
\label{eq41} \phi\sim\left(-\epsilon
\xi_{0}\sqrt{3}\right)^{\frac{1}{2}}t+ \mbox{constant},
\end{equation}
and
\begin{equation}
\label{eq42} V(\phi)\sim\xi_{0}\sqrt{3},
\end{equation}
respectively. Eq. (\ref{eq41}) clearly shows that the only
possible scenario at far future is the phantom scenario as
$\epsilon=1$ (quintessence) gives an imaginary $\phi$. It is worth
to mention that at late time i.e $a\to \infty$ which implies
$\xi_{0}\to 0$, the potential asymptotically tends to vanish and
$\phi=\mbox{constant}$.

\section{Statefinder Diagnostic}
V. Sahni and coworkers (2003) have recently introduced a
pair of parameters $\{r, s\}$ called ``{\it statefinders}", which
are useful to distinguish different types of dark energy. The
statefinders were introduced to characterize primarily flat
universe models with cold dark matter (dust) and dark energy. They
were defined as
\begin{equation}
\label{eq43}
r\equiv\frac{\dot{\ddot{a}}}{aH^{3}},~~~~~s\equiv\frac{r-\Omega}{3(q-\frac{\Omega}{2})}.
\end{equation}
Here the formalism of Sahni and coworkers is extended to permit
curved universe models. If we suppose that dark energy does not
interact with dark matter (as we assumed), then the statefinder
pair can be further expressed as
\begin{equation}
\label{eq44}
r=\Omega_{m}+\frac{9\omega^{de}}{2}\Omega^{de}(1+\omega^{de})-\frac{3}{2}\Omega^{de}\frac{\dot{\omega}^{de}}{H},
\end{equation}
\begin{equation}
\label{eq45}
s=1+\omega^{de}-\frac{1}{3}\frac{\dot{\omega}^{de}}{\omega^{de}H},
\end{equation}
where $\Omega=\Omega_{m}+\Omega_{de}$. The statefinder is a
``geometrical" diagnostic in the sense that it depends upon the
expansion factor and hence upon the metric describing
space-time.\\

If the dark energy is due to a scalar field the
equation of state factor $w^{de}$ is given by
\begin{equation}
\label{eq46}\omega^{de}=\frac{\dot{\phi}^{2}-2\epsilon
V(\phi)}{\dot{\phi}^{2}+2\epsilon V(\phi)}.
\end{equation}
by taking differentiation we get
\begin{equation}
\label{eq47}
\dot{\omega}^{de}\rho^{de}=\frac{2\epsilon\dot{\phi}(2\ddot{\phi}V-\dot{\phi}^{2}\dot{V})}{\dot{\phi}^{2}+2\epsilon
V(\phi)}
\end{equation}
Using the equation of motion for the scalar field
\begin{equation}
\label{eq48} \ddot{\phi}+3H\dot{\phi}+\epsilon V'=0,
\end{equation}
in eq. (\ref{eq47}) and inserting the result into (\ref{eq44}) we
obtain (note that $\dot{V}=V'\dot{\phi},~
V'=\frac{dV(\phi)}{d\phi}$)
\begin{equation}
\label{eq49}
r=\Omega+\frac{3}{2}\frac{\dot{\phi}^{2}}{H^{2}}+\epsilon\frac{\dot{V}}{H^{3}}
\end{equation}
Furthermore, from Raychaudhuri's equation
\begin{equation}
\label{eq50}
 \frac{\ddot{a}}{a} = \frac{3}{2}\xi_{0}H(\rho^{de})^{\tau} -
\frac{1}{6}\rho^{de}(1 + 3\omega^{de}) - \frac{1}{6}\rho^{m}(1 +
3\omega^{m}) - \frac{2}{3}\sigma^{2},
\end{equation}
we find
\begin{equation}
\label{eq51}
q-\frac{\Omega}{2}=\frac{\xi_{0}}{2}H^{2\tau-1}(3\Omega^{de})^{\tau}-\frac{2}{3}\sigma^{2}+\frac{1}{2H^{2}}\left(\frac{1}{2}\epsilon\dot{\phi}^{2}-V\right),
\end{equation}

where $\sigma_{ij}$ is the shear tensor which is given by
\begin{equation}
\label{eq52}\sigma_{ij} = u_{i;j} + \frac{1}{2}(u_{i;k}u^{k}u_{j}
+ u_{j;k}u^{k}u_{i}) + \frac{1}{3}\theta(g_{ij} + u_{i}u_{j}).
\end{equation}

Therefore, the statefinder $s$ is also obtained as
\begin{equation}
\label{eq53}
s=\frac{\dot{\phi}^{2}+\frac{2}{3}\epsilon\frac{\dot{V}}{H}}{\frac{\xi_{0}}{3}H^{3-2\alpha}(3\Omega^{de})^{1-\alpha}-(\frac{2\sigma
H}{\sqrt{3}})^{2}+\left(\frac{1}{2}\epsilon\dot{\phi}^{2}-V\right)}
\end{equation}
To study the behavior of viscous DE more precisely we consider a
toy model in the next section.
\section{Test Model}
To examine our above general results we present a worked example
in this section. For this propose we assume the following scale
factor
\begin{equation}
\label{eq54} a(t)=\sinh(t).
\end{equation}
By assuming a time varying deceleration parameter one can generate
such a scale factor (Amirhashchi et al. 2011). It has also been shown that this
scale factor is stable under metric perturbation (Chen and Kao 2001). In
terms of redshift the above scale factor is
\begin{equation}
\label{eq55} a=\frac{1}{1+z},~~~z=\frac{1}{\sinh(t)}-1.
\end{equation}
In this case one can find the DE energy density $\rho^{de}$, the
bulk viscosity $\xi(\rho^{de})$, deceleration parameter $q$, and
average anisotropy parameter $A_{m}$ as
\[
\rho^{de}=3\coth^{2}(t)+3L\sinh^{-6}(t)-\rho^{m}_{0}\sinh^{-3}(t)
\]
\begin{equation}
\label{eq56}=3\frac{1+(1+z)^{2}}{(1+z)^{4}}+3L(1+z)^{6}-\rho^{m}_{0}(1+z)^{3}
\end{equation}
\[
\xi(\rho^{de})=3\xi_{0}\coth(t)\left[3\coth^{2}(t)+3L\sinh^{-6}(t)-\rho^{m}_{0}\sinh^{-3}(t)\right]^{1-\alpha}
\]
\begin{equation}
\label{eq57}
=3\xi_{0}\frac{\sqrt{1+(1+z)^{2}}}{(1+z)^{2}}\left[3\frac{1+(1+z)^{2}}{(1+z)^{4}}+3L(1+z)^{6}-\rho^{m}_{0}(1+z)^{3}\right]^{1-\alpha}
\end{equation}
\begin{equation}
\label{eq58} q=-\tanh^{2}(t)=-\frac{1}{1+(1+z)^{2}}
\end{equation}
\begin{figure}[ht]
\begin{minipage}[b]{0.5\linewidth}
\centering
\includegraphics[width=\textwidth]{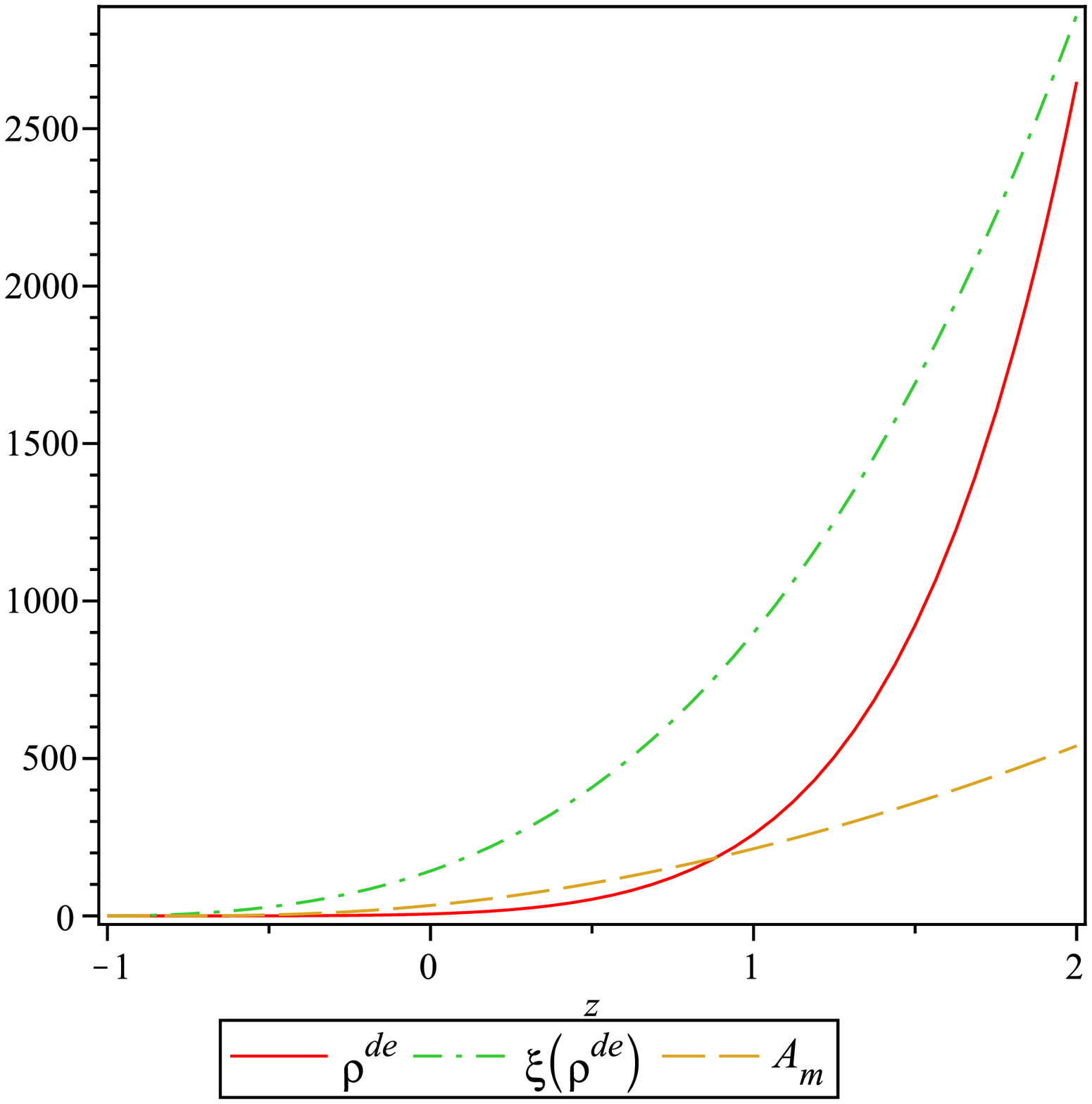} \\
\caption{The plot of the DE energy density $\rho^{de}$, average
anisotropy parameter $A_{m}$, and the bulk viscosity
$\xi(\rho^{de})$ vs. $z$ for $\rho^{m}_{0}=0.24$, $L=0.1$,
$\xi_{0}=0.1$.}
\end{minipage}
\hspace{0.5cm}
\begin{minipage}[b]{0.5\linewidth}
\centering
\includegraphics[width=\textwidth]{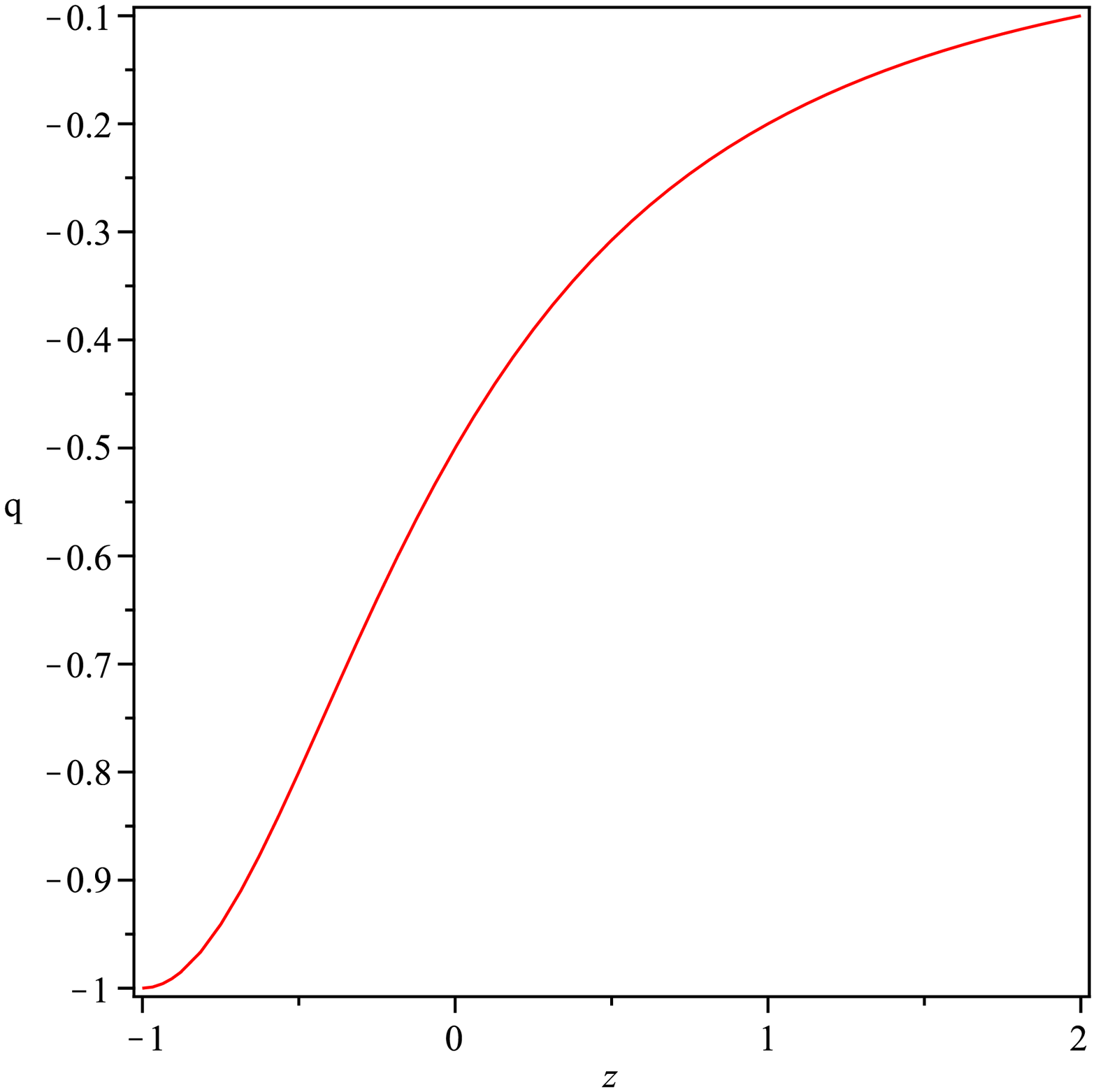}
\caption{ The plot of deceleration parameter $q$ versus redshift
($z$).}
\end{minipage}
\end{figure}
\begin{equation}
\label{59} A_{m}=\frac{1}{3}\sum^{3}_{i}\left(\frac{\triangle
H_{i}}{H}\right)^{2}=\frac{1}{3}(b_{1}^{2}+b_{2}^{2}+b_{3}^{2})\frac{\sinh^{-4}(t)}{1+\sinh^{2}(t)}
=\frac{1}{3}(b_{1}^{2}+b_{2}^{2}+b_{3}^{2})\frac{(1+z)^{4}}{1+(1+z)^{2}}
\end{equation}
where $\triangle H_{i}=H_{i}-H (i=1, 2, 3)$ and
$H_{1}=\frac{\dot{A}}{A}$, $H_{1}=\frac{\dot{B}}{B}$, and
$H_{3}=\frac{\dot{C}}{C}$ are the directional Hubble's parameters
in the directions of $x, y$ and $z$ respectively.\\
Figure $1$ depicts the variation of energy density of $\rho^{de}$,
 $A_{m}$, and $\xi(\rho^{de})$ versus redshift $z$. As it is
 expected all these parameters are decreasing functions and
 approaches to zero at late time ($z=-1$). The variation of
 deceleration parameter (DP) is also shown in Figure $2$. From this
 figure we observe that the value of DP in present time is almost
 $-0.6$ which is in good agreement with the value of DP obtained from
 observations. Also at late time i.e $z=-1$, deceleration
 parameter tends to $-1$ as in the case of de-Sitter universe.\\

By using eq. (\ref{eq54}) in eqs. (\ref{eq34}), (\ref{eq39}), and
(\ref{eq40}) and after simplification the EoS of viscous dark
energy $\omega^{de}_{vf}$, scalar field $\phi$ and the potential
$V(\phi)$ are obtained as
\[
\omega^{de}_{vf}=-\frac{1}{3}\left(\frac{+2\tanh^{2}(t)+1+L\sinh^{-4}(t)\cosh^{-2}(t)}{1+L\sinh^{-4}(t)\cosh^{-2}(t)-\Omega^{m}_{0}\sinh^{-3}(t)}\right)
\]
\begin{equation}
\label{eq60}
-3^{1-\alpha}\xi_{0}\frac{\coth^{1-2\alpha}(t)}{(\Omega^{de})^{\alpha}},
\end{equation}
\begin{equation}
\label{eq61}
\dot{\phi}^{2}=2\epsilon\left[\sinh^{-2}(t)+L\sinh^{-6}(t)-\frac{3}{2}
\Omega^{m}_{0}\sinh^{-5}(t)\cosh^{2}(t)-\frac{\xi_{0}}{2}\sqrt{3\Omega^{de}}\right],
\end{equation}

\begin{equation}
\label{eq62}
V=2\left[2\tanh^{-2}(t)+1-4L\sinh^{-6}(t)-\frac{3}{2}\Omega^{m}_{0}\sinh^{-5}(t)\cosh^{2}(t)+\frac{\xi_{0}}{2}\sqrt{3\Omega^{de}}\right].
\end{equation}
\begin{figure}[ht]
\begin{minipage}[b]{0.5\linewidth}
\centering
\includegraphics[width=\textwidth]{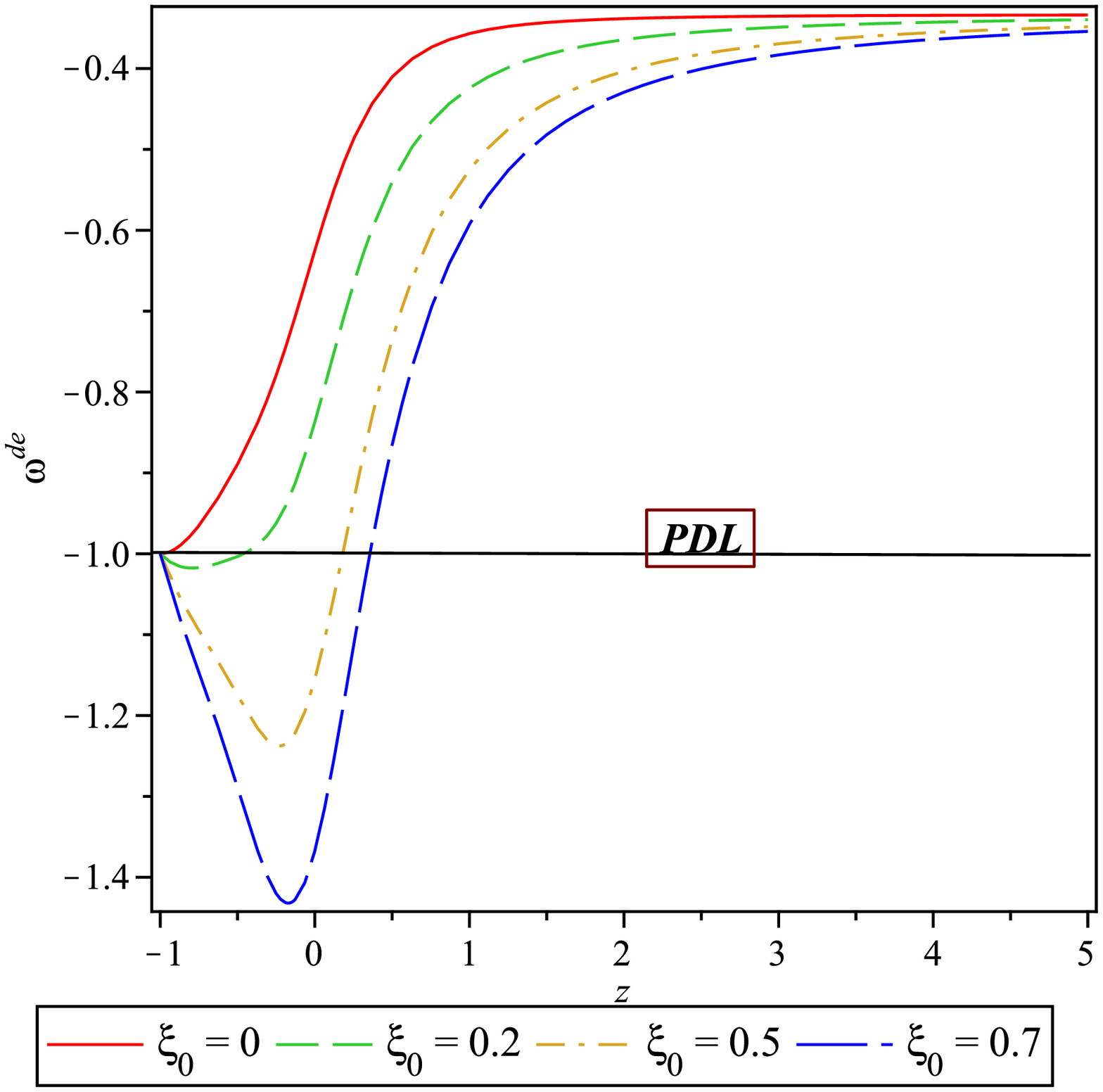} \\
\caption{The plot of EoS parameter versus redshift ($z$) for
$\Omega^{m}_{0}=0.24$, $L=0.1$.}
\end{minipage}
\hspace{0.5cm}
\begin{minipage}[b]{0.5\linewidth}
\centering
\includegraphics[width=\textwidth]{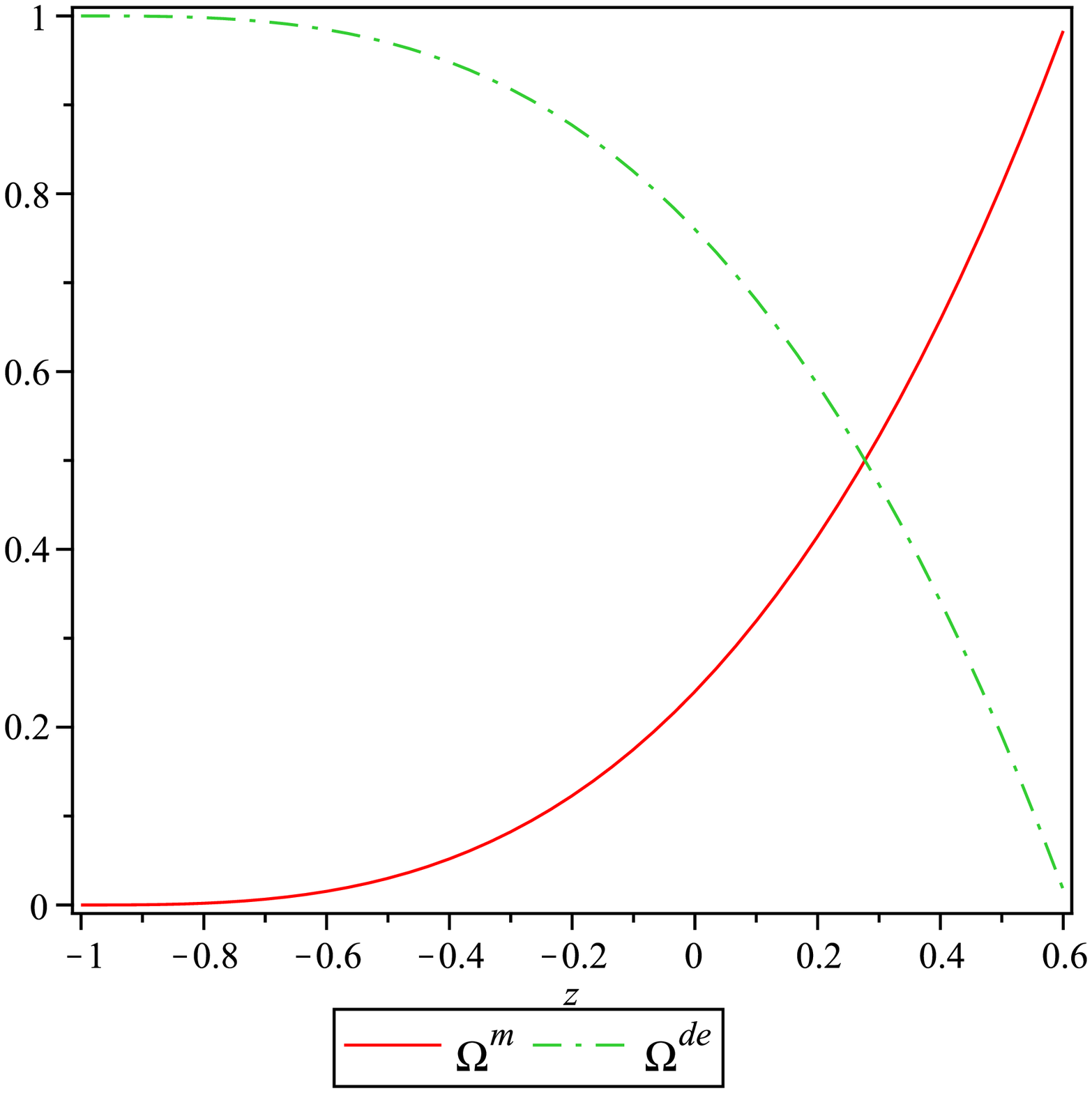}
\caption{ The plot of energy $\Omega^{m}$ and $\Omega^{de}$ versus
redshift ($z$) for $\Omega^{m}_{0}=0.24$, $L=0.1$.}
\end{minipage}
\end{figure}

We can re-write eqs. (\ref{eq60})-(\ref{eq62}) in term of redshift
as
\begin{equation}
\label{eq63}
\omega^{de}_{vf}=-\frac{1}{3}\left(\frac{1+\frac{2}{1+(1+z)^{2}}+
L\frac{(1+z)^{6}}{1+(1+z)^{2}}}{1+L\frac{(1+z)^{6}}{1+(1+z)^{2}}-\Omega^{m}_{0}(1+z)^{-3}}\right)
-3^{1-\alpha}\xi_{0}\frac{[(1+z)^{-4}+(1+z)^{-2}]^{1-2\alpha}}{(\Omega^{de})^{\alpha}},
\end{equation}
\begin{equation}
\label{eq64}
\dot{\phi}^{2}=2\epsilon\left[(1+z)^{2}+L(1+z)^{6}-\frac{3}{2}\Omega^{m}_{0}(1+z)^{3}(1+(1+z)^{2})-\frac{\xi_{0}}{2}\sqrt{3\Omega^{de}}\right],
\end{equation}
\begin{equation}
\label{eq65}
V=2\left[1+\frac{2}{1+(1+z)^{2}}-4L(1+z)^{6}-\frac{3}{2}\Omega^{m}_{0}(1+z)^{3}(1+(1+z)^{2})+\frac{\xi_{0}}{2}\sqrt{3\Omega^{de}}\right].
\end{equation}
The behavior of EoS parameter, $\omega^{de}$, in terms of redshift
$z$ is shown in Fig. $3$. It is observed that the EoS parameter is
a decreasing function of $z$ and the rapidity of its decrease
depends on the value of $\xi_{0}$. We see that in absence of bulk
viscosity the EoS always varying in quintessence region (red
line/solid line) whereas in presence of viscosity EoS cross PDL
and varying in phantom region. But at the later stage of evolution
it tends to the same constant value i.e $\omega^{de}=-1$
independent of the value of $\xi_{0}$. This behavior clearly shows
that the phantom phase i.e $\omega^{de}<-1$ is an unstable phase
and there is a transition from phantom to the cosmological
constant phase at late time. As we mention above, the phantom
phase instability of the universe is because of the fact that the
viscosity dies out as time is passing.\\

The matter density $\Omega^{m}$ and dark energy density
$\Omega^{de}$ can be easily calculated as
\begin{equation}
\label{eq66}
\Omega^{m}=\Omega^{m}_{0}\sinh^{-3}(t)=\Omega^{m}_{0}(1+z)^{3},
\end{equation}
\begin{equation}
\label{eq67}
\Omega^{de}=1+L\sinh^{-4}(t)\cosh^{-2}(t)-\Omega^{m}_{0}\sinh^{-3}(t)=1+L\frac{(1+z)^{6}}{1+(1+z)^{2}}-\Omega^{m}_{0}(1+z)^{3}.
\end{equation}
Also from above two equations we obtain the total energy density
as
\begin{equation}
\label{eq68}
\Omega=\Omega^{m}+\Omega^{de}=1+L\frac{(1+z)^{6}}{1+(1+z)^{2}}
\end{equation}

The variation of density parameters $\Omega^{m}$ and $\Omega^{de}$
with redshift $z$ have been shown in Fig. $4$. Here, we observe
that $\Omega^{de}$ increases as redshift decreases and approaches
to $1$ at late time whereas
$\Omega^{m}$ decreases as $z$ decreases and approaches to zero at late time.\\
\begin{figure}[ht]
\begin{minipage}[b]{0.5\linewidth}
\centering
\includegraphics[width=\textwidth]{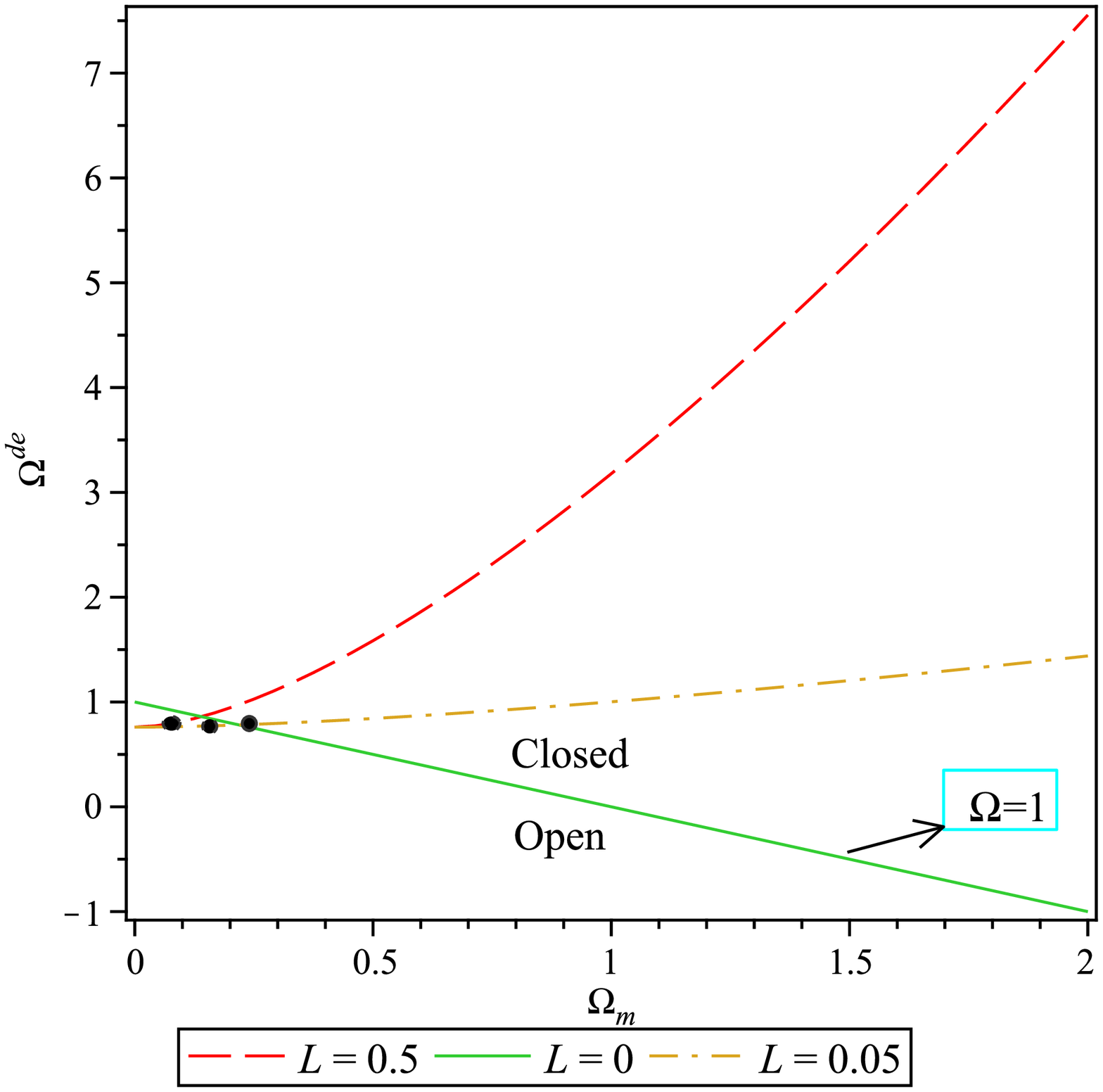} \\
\caption{The plot of $\Omega^{de}$ versus $\Omega^{m}$. The solid
line indicates flat universe (L=0). The dots locate the current
values of $\Omega^{de}$ and $\Omega^{m}$ for $L=0,~ 0.05,~ 0.5$}
\end{minipage}
\hspace{0.5cm}
\begin{minipage}[b]{0.5\linewidth}
\centering
\includegraphics[width=\textwidth]{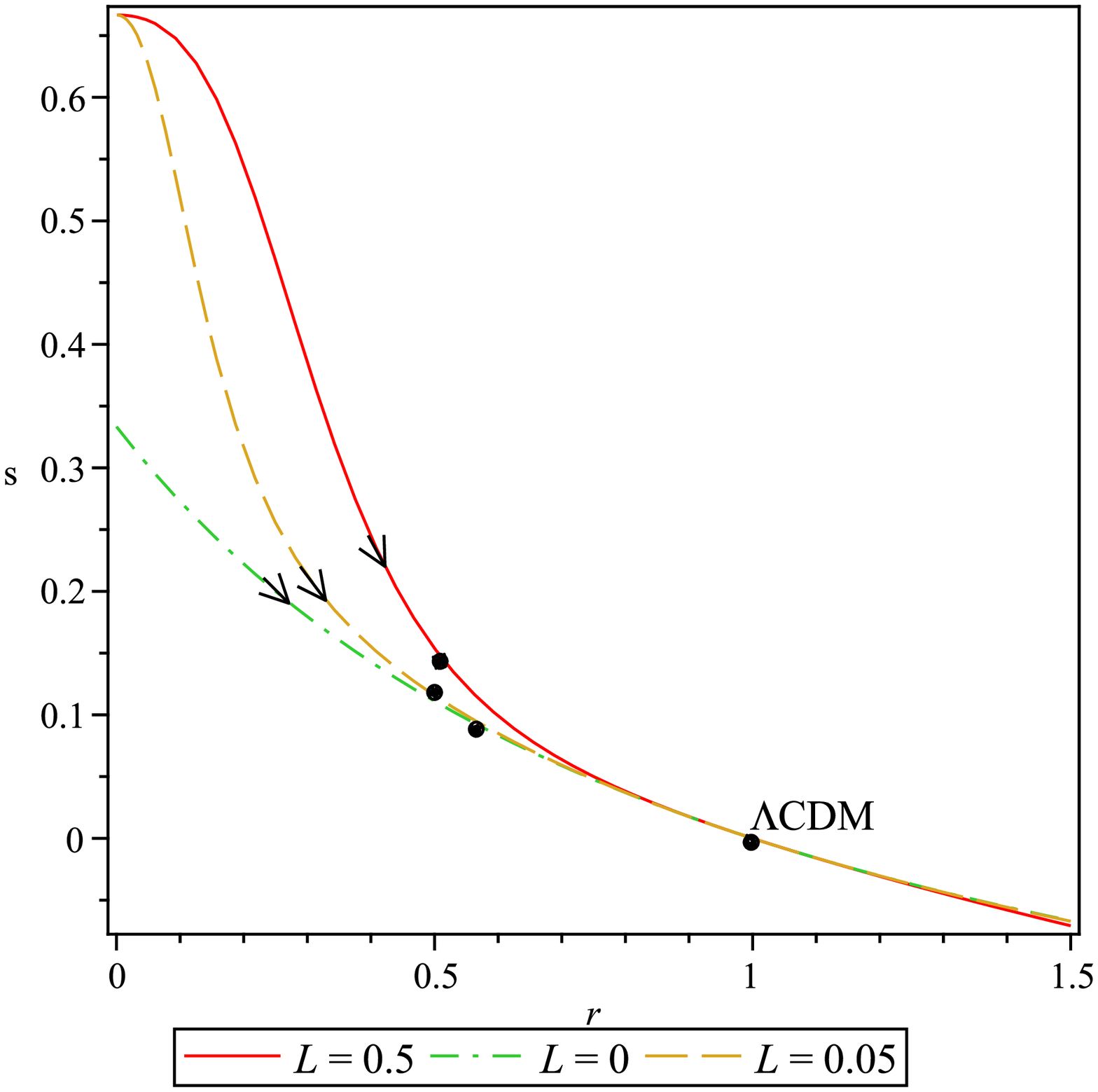}
\caption{ $s-r$ evolution diagram. The dots locate the current
values of the statefinder pair $\{r, s\}$.}
\end{minipage}
\end{figure}
For our model, the parameters $\{r, s\}$ can be explicitly written
in terms of cosmic time $t$ or redsfift $z$ as
\begin{equation}
\label{eq69} r=\tanh^{2}(t)=\frac{1}{1+(1+z)^{2}}
\end{equation}
and
\begin{equation}
\label{eq70}
s=\frac{1+L\sinh^{-4}(t)\cosh^{-2}(t)-\tanh^{2}(t)}{\frac{3}{2}\left[1+2\tanh^{2}(t)+L\sinh^{-4}(t)\cosh^{-2}(t)\right]}=
\frac{(1+z)^{2}\left(1+L(1+z)^{4}\right)}{\frac{3}{2}\left[2+L(1+z)^{2}\left(1+(1+z)^{4}\right)\right]}
\end{equation}
Figure $5$ shows the values of $\Omega^{de}_{0}$ and
$\Omega^{m}_{0}$ which are permitted by our model. From this
figure we observe that for case $L=0$ which represents a spatially
flat universe ($\Omega=1$), $\Omega^{de}_{0}\approx 0.76$ and
$\Omega^{m}_{0}\approx 0.24$. These results are in good agreement
with the CMB results, the supernova results, and the computed
density of matter in clusters. Other models with $L\neq 0$,
represent open universes with $\Omega<1$.\\ Trajectories in $s-r$
plane corresponding to different cosmological models are shown in
figure $6$. The dots in the diagram locate the current values of
the statefinder pairs $\{s,r\}$. From this figure we see
explicitly that the ingredient parameter $L$ (or $K$) makes the
model evolve along different trajectories on the $s-r$ plane. It
is worth to mention that the cold dark matter with a cosmological
constant ($\Lambda CDM$) diagrams (spatially flat) corresponds to
the fixed point $\{s,r\}_{\Lambda CDM}=\{0,1\}$. From eqs.
(\ref{eq69}) and (\ref{eq70}) we obviously see that
$\{s,r\}=\{0,1\}$ at late time i.e $z=-1$.
\section{Concluding Remarks}
Phantom field models have been suggested in order to provide a
theoretical support for the recent observation that mildly favor
the EoS of DE crossing $-1$ near the past. A lot of studies have
been done in this regard and many phantom field models have been
proposed. Some of these models are evolving from quintessence to
phantom called quintom. However, theses models suffer from two
major problems i.e. (1) Instability of phantom field and (2)
finite future singularity (big rip). In this paper we proposed a
simple mechanism to alleviate these problems by introducing a
special form of bulk viscosity i. e. $\Pi = -3\xi_{0}H(\rho^{de})^{\tau}$
in the cosmic fluid. In this
mechanism first, viscosity causes dark energy which is varying in
quintessence to pass phantom divided line (PDL) and drop it to the
phantom region but since viscosity is a decreasing function of
time, as time is passing it dies out and $\omega^{de}$ leaves
phantom region and tends to $-1$ at late time. Hence the problem
of future singularity (big rip) does not occur in this scenario.
To test the impact of the anisotropy parameter ($L$), we perform a
statefinder diagnostic on this scenario. This diagnostic shows
that the statefinder parameters can probe the anisotropy of the
model. May be future SNAP would be capable of probing this effect.
In summary, The general form of the EoS parameter of viscous and
non-viscous dark energy has been investigated in this paper. It is
found that the presence of bulk viscosity causes our universe to
get to the darker region i.e phantom temporarily. It is worth to mention that since our anisotropic model behaves as isotropic FLRW universe at late time, as a result, the phantom does not survive in isotropic universe as well. Our results fulfil the theoretical requirement argued by Carroll et all (2003) which state that, to avoid the big rib problem, all phantom models should decay to cosmological constant at late time. Moreover, since we have not restricted our study to the maximally symmetric FLRW space-times, our results seems to be more general than those obtained on the bases of this isotropic universes.
\section*{Acknowledgments}
Author would like to thank Laboratory of Computational Sciences
and Mathematical Physics, Institute for Mathematical Research,
Universiti Putra Malaysia for providing facility where this work
was done. Author also would like to acknowledge the anonymous
referee for fruitful comments.

\end{document}